\documentclass[12pt,preprint]{aastex}
\shorttitle{Masses of nuclear clusters in bulge-less galaxies}
\shortauthors{Walcher et al.}
\begin{document}
\title{Masses of star clusters in the nuclei of bulge-less spiral galaxies}

\author{C.~J. Walcher\altaffilmark{1}, 
R.~P. van der Marel\altaffilmark{2}, 
D. McLaughlin\altaffilmark{2},
H.-W. Rix\altaffilmark{1},
T. B\"oker\altaffilmark{3},
N. H\"aring\altaffilmark{1},
L.~C. Ho\altaffilmark{4},
M. Sarzi\altaffilmark{5},
J.~C. Shields\altaffilmark{6}}

\altaffiltext{1}{Max Planck Institut f\"ur Astronomie, K\"onigstuhl 17, 
D-69117 Heidelberg, Germany}
\altaffiltext{2}{Space Telescope Science Institute, 3700 San Martin
Drive, Baltimore, MD 21218}
\altaffiltext{3}{ESA/ESTEC, Keplerlaan 1, 2200 AG Noordwijk, Netherlands}
\altaffiltext{4}{The Observatories of the Carnegie Institution of Washington, 813 Santa Barbara Street, Pasadena, CA 91101-1292}
\altaffiltext{5}{Oxford Astrophysics, Keble Road, Oxford, OX13RH, UK}
\altaffiltext{6}{Department of Physics and Astronomy, Clippinger Research Laboratories, Ohio University, 251B, Athens, OH45701-2979}

\begin{abstract}

In the last decade star clusters have been found 
in the centers of spiral galaxies across all Hubble types. 
We here present a spectroscopic study of the exceptionally 
bright ($10^6$ -- $10^8$ L$_{\odot}$) but compact 
($r_e \sim$ 5 pc) nuclear star clusters in 
very late type spirals with UVES at the VLT. We find the 
velocity dispersions of the nine clusters 
in our sample to range from 13 to 34 km~s$^{-1}$. Using 
photometric data from the HST/WFPC2 and spherically symmetric 
dynamical models we determine masses between $8 \times 10^5$ and 
$6 \times 10^7$ M$_\sun$. 
The mass to light ratios range from 0.2 to 1.5 in the I band 
This indicates a young mean age for most clusters, in agreement with 
previous studies. 
Given their high masses and small sizes we find that nuclear clusters 
are among the objects with the highest mean surface density known 
(up to $10^5$ M$_{\sun}$ pc$^{-2}$).
From their dynamical properties we infer that, rather than small bulges, 
the closest structural kin of 
nuclear clusters appear to be massive compact star clusters. 
This includes such different objects as globular clusters, 
"super star clusters", ultra compact dwarf galaxies and the nuclei 
of dwarf elliptical galaxies. It is a challenge to explain 
why, despite the wildly different current environments, all 
different types of massive star clusters share very similar 
and structural properties. 
A possible explanation links UCDs and 
massive globular clusters to nuclear star clusters through 
stripping of nucleated dwarf galaxies in a merger event. 
The extreme properties of this type of clusters would then be 
a consequence of their location in the centers 
of their respective host galaxies.

\end{abstract}


%%%%%%%%%%%%%%%%%%%%%%%%%%%%%%%%%%%%%%%%%%%%%%%%%%%%%%%%%%%%%
\section{Introduction}
\label{s:intro} 

In the last decade the image quality of HST has boosted the study of the 
centers of spiral galaxies. These observations have shown that star clusters 
are a common feature in the nuclei of spiral galaxies of all Hubble types 
(Phillips et al. 1996; Matthews \& Gallagher 1997; Carollo et al. 1997;  
Carollo, Stiavelli \& Mack 1998; 
B\"oker et al. 2002), from bulge-dominated galaxies to bulge-less galaxies. 
Nuclear star clusters are hard to observe in early type spirals against the 
bright bulge. However in late-type, bulge-less galaxies, they stand out well 
against the low surface brightness disk. The HST survey by B\"oker et al. 
(2002, hereafter B02) has shown that 75\% of all late type spirals host such a 
nuclear cluster  in their photometric center. We use the term 
nuclear cluster (NC) for a luminous and compact star cluster near 
the overall photometric center of the galaxy. While the effective 
radii ($\sim$ 5 pc, B\"oker et al. 2004, 
hereafter B04) are comparable to globular clusters, the luminosities 
exceed those of the most luminous Milky Way globular clusters by up to 2 
orders of magnitude. 

In the absence of any obvious bulge 
(B\"oker, Stanek \& van der Marel 2003), NCs represent the closest 
thing to a central ``hot component''. 
In most galaxy formation scenarios the bulge at the 
center is the ``trashbin of violent relaxation'', where a kinematically 
hot stellar component has formed either through external potential 
perturbations, such as early mergers of fragments, or perhaps
through internal effects such as violent bar instabilities. 
Bulge-less galaxies in contrast, must have lived very sheltered 
lives, as their central ``trashbin'' is virtually empty.
The discovery of almost ubiquitous NCs at the same physical location 
in their host galaxy as a bulge would have, brings up the question 
whether bulges and NCs are related. 
A thinkable scenario would be that NCs are protobulges that 
grow by repeated accretion of gas and subsequent star formation. 
The pace of that growth and the resulting size of the central component 
would then determine if we call the central component a NC or a small bulge. 

While the high spatial resolution of HST allows surface photometry 
of the clusters, other fundamental parameters such as e.g. age and mass 
remain largely unknown. Several case studies in the literature however have 
revealed massive, young objects. 
M33 is the nearest Sc galaxy hosting a NC and its nucleus has been 
extensively studied in the past decade (e.g. Davidge 2000; Long, Charles 
\& Dubus 2002; Stephens \& Frogel 2002). Despite some differences in the 
details, all studies on M33 agree that there is some population younger than 
0.5 Gyr in the central parsec and that star formation has varied significantly 
over the past several Gyrs. The mass of the central cluster in M33 was 
estimated from a detailed population analysis in Gordon et al. (1999) to 
be $5\times 10^5$ M$_{\sun}$, consistent with the upper limit derived 
from the velocity dispersion by Kormendy \& McClure (1993) of $2\times 
10^6$ M$_{\sun}$.
From H$\alpha$ rotation curves of late type spirals, Matthews \& Gallagher 
(2002) find that the velocity offsets at the position of five semi-stellar
nuclei --- certainly to be identified with NCs --- 
are consistent with masses of $\approx 10^6$ -- $10^7$ M$_{\sun}$. 
They also point out that the location of the cluster and the 
dynamical center of the galaxy do not always coincide. The only direct mass 
determination for a NC from a measurement of the stellar velocity 
dispersion and detailed dynamical modeling was done for the NC in IC342
by B\"oker, van der Marel \& Vacca (1999); they find 
$6\times 10^6$ M$_{\sun}$, with a K-band mass-to-light ratio of 0.05 
in solar units. The derived mean age is $\le 10^8$ 
years. B\"oker et al. (2001) also studied the NC in NGC4449 using population 
synthesis models. They estimate an age of $\approx 10^7$ years 
in agreement with Gelatt, Hunter \& Gallagher (2001). They infer a lower 
limit for the mass, which is $4\times 10^5$ M$_{\sun}$.

The available sizes, masses and ages suggest kinship between NCs and 
either globular clusters (GCs), super star clusters (SSCs) or the nuclei 
of dwarf ellipticals (dE). Alternatively, NCs could be a new class of 
objects which would then be part of the wide range of special phenomena 
occuring in galaxy centers. To establish the relation between these objects, 
we clearly need reliable 
determinations of the velocity dispersion $\sigma$ and therefore masses 
as well as ages for a statistically significant sample of objects. 
This is the first paper in a series of two describing the results of 
ground-based follow-up spectroscopy with the Ultraviolet and Visual Echelle 
Spectrograph (UVES) at the VLT 
to the HST survey by B\"oker et al. (2002). Here we focus on the masses 
and velocity dispersions of the NCs, while a second paper 
(Walcher et al. 2004, in preparation) will deal with ages and stellar 
populations.

%%%%%%%%%%%%%%%%%%%%%%%%%%%%%%%%%%%%%%%%%%%%%%%%%%%%%%%%%%%%%
\section{Sample selection and data}
\label{s:data} 

\subsection{Sample selection}
All objects were drawn from the sample in B02. 
Galaxies in this survey were selected to have Hubble type between 
Scd and Sm, to have line-of-sight velocity $v_{\mbox{hel}} < 2000 $km~s$^{-1}$ 
and to be nearly face-on. A subset of 9 objects (listed in Table 
\ref{t:UVESsample}) were observed with UVES at 
the VLT. The objects were selected from the full catalogue to be 
accessible on the sky and to be bright 
enough to be observed in less than three hours, maximizing the 
number of observable objects. We thus sample the brighter 2/3 of the 
luminosity range covered by the clusters identified from the HST images. 
Whether this bias in absolute 
magnitude introduces others, either towards younger or more massive 
clusters, is at present unclear.

\subsection{Observations}
The VLT spectra were taken with the Ultraviolet and Visual Echelle 
Spectrograph (UVES) attached to Kueyen (UT2) during three nights of 
observation 
from the 9th to the 11th of December 2001. All nights were clear with a 
seeing around 0.8$^{\prime\prime}$. UVES can work in parallel at blue 
and red wavelengths by using a dichroic beam splitter that divides the 
light from the object at the entrance of the instrument. Our main goal 
was to derive ages from the spectral region around the Balmer break and 
velocity dispersions from the Calcium Triplet. In order to also include 
the H$\alpha$ line we chose a 
non-standard setting with the dichroic \#2 and the crossdispersers \#2 
and \#4 centered on the wavelengths 4200 and 8000 {\AA}. The resulting 
wavelength range is thus 3570 -- 4830 {\AA} in the blue arm; the red arm 
spectra are imaged on two CCDs, one covering 6120 -- 7980 {\AA}  and the 
other one covering 8070 -- 9920 {\AA}. The length of the slit was 
10$^{\prime\prime}$, the width 1$^{\prime\prime}$. The slit 
was always oriented perpendicular to the horizon to minimize effects of 
differential refraction. The spectral resolution is $\approx$35000 
around the Calcium Triplet as measured from sky lines. This corresponds 
to a Gaussian dispersion of the instrumental line-spread function of 3.4 
km~s$^{-1}$. The pixel size in the reduced spectra is $\sim 0.07$ \AA.

Effective radii $r_e$ for the clusters were derived from HST photometry in 
B04. They are smaller than 0.2$^{\prime\prime}$ for all observed clusters.
Because the seeing disk is bigger than the effective radii we essentially 
measure integrated properties. The observed NCs together with exposure 
times are listed in Table \ref{t:UVESsample}. For all objects in our sample 
the HST images can be found in B02.

We also observed a number of template stars of known spectral type ranging 
from B to K and listed in table \ref{t:UVESstars}. As will be explained in 
Section \ref{s:veldisp} these are used to reliably measure the velocity 
dispersion of the NCs. 

\subsection{Reduction}
\label{s:reduction}

The data were reduced with the UVES reduction pipeline version 1.2.0 
provided by ESO (see ``UVES pipeline and Quality Control User's Manual'' 
prepared by P.Ballester et al., 2001 and the website: 
http://\-www.eso.org/\-instruments/\-uves/). For each observing night the 
following steps for preparing the calibration database were performed: 
first, using a physical model of UVES a first guess wavelength 
solution was determined. Then the order positions were identified from an 
order definition flatfield. Next the final wavelength calibration was done 
using a ThAr-lamp exposure. Finally master bias and master flatfields 
were created from the median of five input frames each.

Cosmic ray removal was done on the 2-D science frames by means of the MIDAS 
routine {\tt filter/cosmic}. The science frames were then bias subtracted and 
the interorder background was subtracted by fitting a spline function to the 
grid of background positions. For each background position the median of 
pixels within a certain window was used as the measurement at that point. The 
flatfielding was not done on the 2D raw frames, but after extraction for each 
order on the extracted 1D spectrum. 
The extraction can be done using either an optimal 
extraction method or a simple average over a predefined slit. For our 
high signal-to-noise (S/N) data 
we found that the optimal extraction method shows quality problems appearing 
as sudden spikes and ripples, especially in the blue. We were able to recover 
almost the same S/N with the average method also for the lower S/N spectra by 
fine-tuning the extraction parameters (position and width of the extraction 
slit and position of the sky windows). For consistency we therefore decided to 
extract all spectra with the average method. S/N values around the Calcium Triplet 
are listed for all spectra in Table \ref{t:UVESsample}. 

It is important to note that background light from the galaxy disk 
underlying the NC is subtracted together with the sky background during 
extraction, so that the contamination from non-cluster light is reduced. 
However, not all the disk light is subtracted, because the disk brightness 
is not generally constant with radius. To quantify the fraction of 
cluster light in the final spectra we simulate our UVES observations 
using the high spatial resolution HST data. The I-band, at an 
effective wavelength of 8000 {\AA}, will yield a correct fraction 
for the wavelength of the Ca Triplet.
We convolve the HST images with a Gaussian of 
0.8$^{\prime\prime}$ seeing. From the known slit position we can derive the 
position of the spectroscopic object and sky windows on the HST image. 
The expected total flux in the reduced object 
spectrum $F_{\rm spec}$ is then computed by subtracting the flux 
in the sky windows from that in the object window. The total flux F$_{\rm cl}$ 
from the cluster alone is known from the magnitudes given 
in B02. The fraction $f$ of the total cluster flux  that falls within the 
spectroscopic object window is 
\begin{equation}
  {\rm f} = \frac{1}{4} [{\rm erf}( x/ \sqrt{2} \sigma)-{\rm erf}(- x/ \sqrt{2} \sigma) ] [{\rm erf}( y/ \sqrt{2} \sigma)-{\rm erf}(- y/ \sqrt{2} \sigma) ]
\label{lightfrac}
\end{equation}
where 
\begin{equation}
  \sigma = \sqrt{\sigma_{\rm PSF}^2 + \sigma_{\rm cluster}^2 }.
\end{equation}
Here $x$ is half the size of the spectroscopic object extraction window 
along the slit and $y$ is half the slit width. In this equation it is 
assumed for simplicity that the PSF and cluster are well approximated 
by gaussians with the correct FWHM. Finally the contamination from 
non-cluster light (NCL) in the extracted spectra 
(as given in Table \ref{t:UVESsample}) is: 
\begin{equation}
 {\rm NCL} =  1 - \frac{ f \times F_{\rm cl}}{  F_{\rm spec}}.
\end{equation}

%%%%%%%%%%%%%%%%%%%%%%%%%%%%%%%%%%%%%%%%%%%%%%%%%%%%%%%%%%%%%
\section{Data analysis}
\label{s:analysis}

\subsection{Measurement of the velocity dispersion}
\label{s:veldisp}

Stellar kinematical analyses of unresolved populations 
assume that an observed galaxy spectrum can be 
represented by the convolution of one or several stellar templates with 
a certain broadening function, often approximated as a Gaussian. 
The linewidth in the template spectra represents both the intrinsic width of 
the stellar photospheric lines and the instrumental broadening. The galaxy 
spectra of unresolved populations are presumed to only differ by the 
additional Doppler broadening, reflecting the velocity dispersion of the 
stars. For massive galaxies this last broadening dominates the intrinsic 
linewidth by a wide margin, and the estimated velocity dispersion is 
nearly insensitive to the choice of templates. This is however not the case 
for the objects in our sample as we will show that their measured velocity 
dispersions range from only 13 to 34 km~s$^{-1}$. 

We use a code that matches simultaneously a linear combination of 
template stars and the width of a single Gaussian broadening function. 
No Fourier transform is involved, as the code works directly with 
the measured spectra (see Rix \& White 1992). Each pixel in the spectrum 
is weighted with its specific error during the fitting process. 
It is particularly important to take templates observed with the same 
instrumental setup as the science targets to avoid systematic errors, 
i.e. resolution mismatch interpreted as velocity dispersion differences. 
We use the seven template stars ranging from type B to K listed in Table 
\ref{s:data}. An inherent 
strength of the code is the ability to determine an optimized template 
from the linear combination of the available template stars. It is thus 
possible to account to some extent for the influence of the generally young 
mean age of the clusters on the intrinsic width of the absorption lines.
A further advantage of the direct fitting in pixel space is that 
masking spectral regions, which is essential to our analysis, is 
straightfoward. 

Velocity dispersions are measured from the spectral region around the 
Calcium Triplet. We choose the largest region that is relatively free 
of telluric absorption bands and emission lines (8400 -- 8900 {\AA}, 
see below). The Calcium Triplet region does 
not only include the very prominent Ca \textsc{II} lines, but also 
several other metal lines whose width can be measured at our 
signal-to-noise ratio. Other regions 
of common use to measure velocity dispersions (as e.g. the CO bandhead 
in the infrared, see e.g. B\"oker et al. (1999) or the region 
around 5100-5500 {\AA} 
as in Maraston et al. (2004)) are not covered by our data. 
We however also investigated two other regions of the observed 
spectrum, namely 6385 -- 6525 {\AA} and 4380 -- 4800 {\AA} which are also free of telluric 
lines and do have other metal lines. The measured velocity dispersions agree to 15 \% 
for those clusters where the signal-to-noise ratio of the metal lines 
is sufficient to measure $\sigma$ reliably. These are most notably the 
old clusters NGC300 and NGC428 and the high 
signal-to-noise spectrum of NGC7793. 
The Calcium Triplet has 
been shown to give reliable results for old 
stellar populations, e.g. in elliptical galaxies, by a variety of 
authors (e.g. Dressler 1984; Barth, Ho \& Sargent 2002). However, as the 
analysis of the blue parts of the UVES spectra show (Walcher et al. 2003), 
some of our objects have ages between $10^7$ and $10^8$ 
years. Unfortunately, kinematic measurements are more 
complicated in populations 
significantly younger than 1 Gyr. Although the intrinsic width of the Ca 
triplet changes only slightly from main-sequence K to late F stars, for 
earlier types rotation and temperature cause the intrinsic width to 
increase rapidly. Further, in supergiants the damping wings of the Voigt 
profile become increasingly important due to the depth of the lines. 
It is in this context very unfortunate that we do not have a supergiant 
in our stellar template set. 
A further complication comes from the presence of intrinsically 
broad absorption lines from the Paschen 
series. Especially the lines Pa16, Pa17 and Pa13 (8504.8, 8547.7 and 
8667.9 {\AA}, respectively) overlap with the Ca-Triplet lines. 
All of these effects lead to an overestimation of the velocity 
dispersion, if only measured from main sequence stars. We thus need 
to analyze carefully the composition of the stellar populations in 
consideration. On the other hand the intervening Paschen absorption 
lines provide an additional diagnostic tool for the age of the 
population. With good signal-to-noise in the spectra this can be used to 
constrain the optimal stellar template mix for the velocity dispersion 
determination. For all spectra we also have age information from the blue 
parts of the spectrum. While we use this information for assessment 
of age related problems, a more detailed analysis of the age composition 
of the clusters will be published in a forthcoming paper (Walcher et al. 
2004 in prep.).

\begin{figure}[t]
\plotone{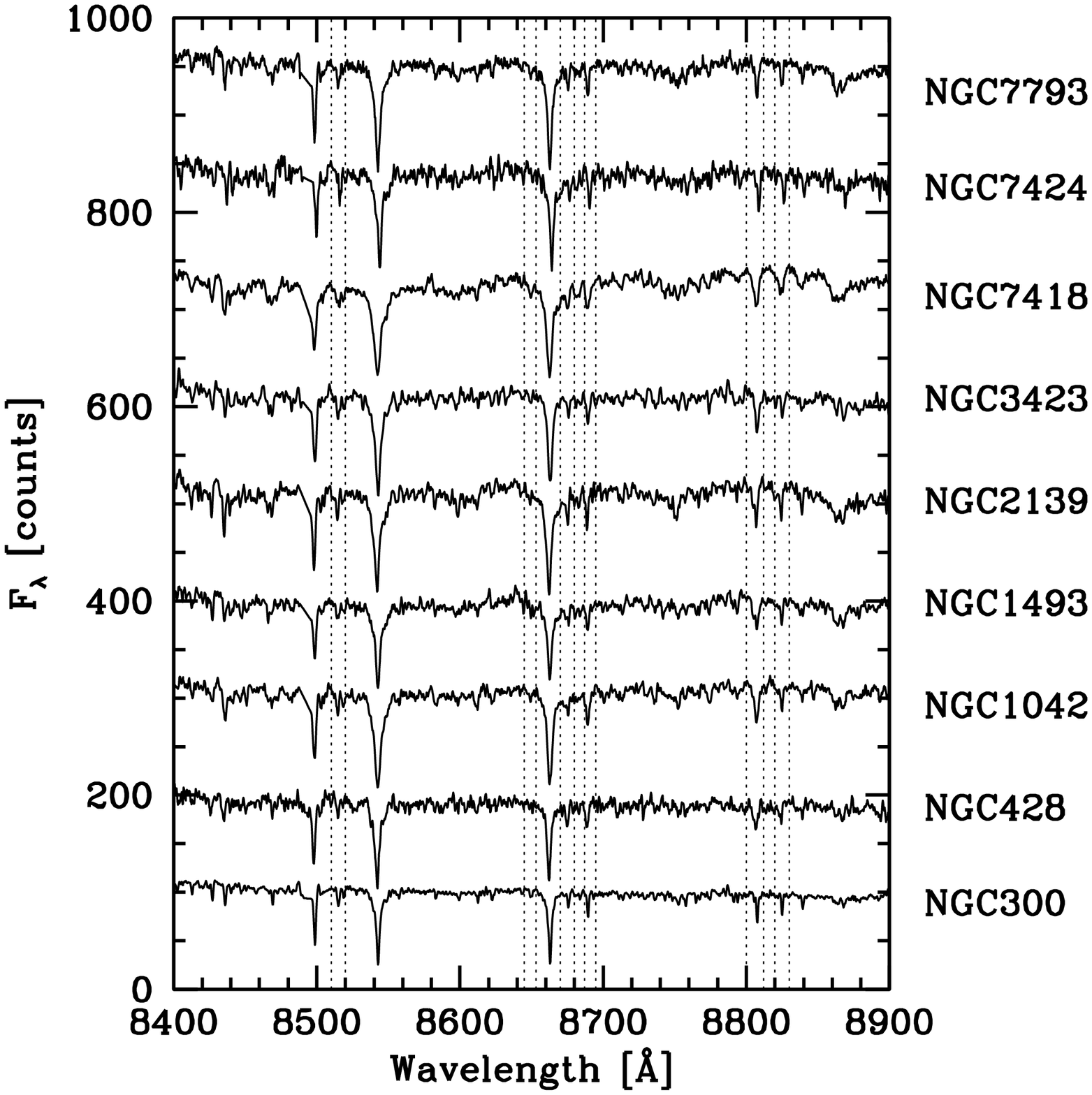}
\caption[NGC1042]{The region around the Calcium Triplet used for the 
velocity dispersion measurements. For better presentation, the spectra 
have been clipped to exclude residuals from sky lines and binned 
over three pixels. Fluxes have been arbitrarily adjusted to fit into 
the plot. Almost 
the full region was used for case 1 fitting, excluding only the region 
from 8470 to 8510 {\AA} because of a CCD defect. The dotted lines show the 
six metal lines used for case 2 fitting.}
\label{f:catripall}
\end{figure}

To assess the influence of the Calcium Triplet problems we follow two 
different 
approaches concerning the wavelength coverage of the velocity dispersion fits.
In what we will call ``case 1'' we use almost the full wavelength range 
between 8400 and 8900 {\AA} as the fitting region. Unfortunately, the 
Ca-Triplet line at 8498 {\AA} falls onto a CCD defect in the interval 
from 8470 -- 8510 {\AA}, which we exclude from all fits. 
We thus cover the two Ca lines 
at 8542 and 8662 {\AA} and further metal lines, mostly Fe \textsc{I}. For case 1, 
the $\chi^2$ that measures the quality of the fit 
is dominated by the Calcium Triplet lines. In ``case 2'' we use only the 
weaker metal lines of Fe and Si in the region for the fit. These are not 
as deep as the Ca \textsc{II} lines and tend to disappear in earlier type stellar 
spectra, so that they are much less affected by the systematic problems 
with the Ca triplet mentioned above. Figure \ref{f:catripall} illustrates 
the spectral region in use. Case 2 clearly is limited by S/N in the galaxy 
spectra and increases possible systematic uncertainties because of 
metallicity mismatches. 
So we divide our sample in two bins, the ``old'' ($\gtrsim 1$ Gyr) 
and the ``young'' ($< 1$ Gyr) bin. The three clusters NGC300, NGC428 and 
NGC3423 are old and we therefore use the case 1 value because 
of the larger fitting 
region and thus better constrained fit. In the other 6 clusters, the young 
clusters, we prefer the case 2 values because they  
avoid the systematic uncertainties related with the Ca triplet. 
For all cases the velocity dispersion for case 1 is higher than 
for case 2. However the mean offset between case 1 and case 2 is 
4.3 km~s$^{-1}$ in the old bin, but 
11.6 km~s$^{-1}$ in the young bin, demonstrating the influence of age 
on the velocity dispersion as measured from the Ca triplet. Table 
\ref{t:UVESstars} quotes the weights that the fitting code 
attributes to every single template star in its best fitting composite 
template. Note that the division in 
an old and a young bin is born out by the lower contribution of A stars 
to the three NCs NGC300, NGC428 and NGC3423. 
Table \ref{t:masses} contains only the finally adopted velocity dispersion.
Formal random errors are available from $\chi ^2$ statistics. However, 
these do not include any systematic 
uncertainties and are therefore not particularly realistic. 
Useful error estimates are hard to obtain with all the complications 
from systematic effects just discussed. 
Typical systematic uncertainties from fitting different initial 
templates and different subsets of the available template set, are of the 
order of 15\%. These are the errors we quote in Table \ref{t:masses}.

The measured velocity dispersion is the luminosity weighted 
mean of the two components (disk and nuclear cluster) that 
contribute to the light inside the ground-based aperture. 
Section \ref{s:masses} details how we account for this in 
our dynamical modelling. Typical disk velocity 
dispersions are around 30 km/s for early type spirals 
(Bottema 1993). The Sm galaxy UGC4325 has a central disk 
velocity dispersion of 19+-2km/s (Swaters 1999). 
Velocity dispersion of cluster and disk 
could therefore be of comparable magnitude.

\subsection{Surface brightness profiles}
\label{s:SFprof}

For the purposes of dynamical modeling we need PSF-deconvolved surface 
brightness profiles of the nuclear clusters and the surrounding galaxy 
disk. To this end we derive I-band surface brightness profiles from 
the HST/WFPC2 imaging data described in B02. Data reduction has been described 
there and will not be repeated here. While B02 inferred surface brightness 
profiles from their data, they did not perform PSF deconvolution. However 
accurate effective radii for most nuclear clusters in the B02 sample have been 
measured for most objects in our sample in B04. For their analysis, B04 
used the software package {\tt ISHAPE} 
(Larsen 1999) to fit PSF-convolved King or Moffat profiles with different 
concentration indices to the data. We here present 
a new analysis with the {\tt galfit} routine (Peng et al. 2002). The reasons 
to repeat the excercise with a different software are twofold: 

\begin{itemize}
\item Three out of the nine clusters in our present sample are placed in 
galaxy centers with irregular morphologies. As the fits from {\tt ISHAPE} were 
therefore deemed unsatisfactory in B04, they were rejected 
at the time. A careful reexamination with {\tt galfit} however yielded useful 
results at least for the purposes of our dynamical modeling.

\item In B04 the main goal was to derive 
accurate effective radii for the clusters alone, implying the use 
of a single component for the cluster profile only. This one component 
then does not represent the light profile outside the cluster radius, 
called R$_u$ in B04. However we here need the 
full deconvolved surface brightness profile, 
including the background galaxy, for the dynamical modeling. {\tt Galfit} 
allows us to model the cluster and its surroundings with an unlimited 
number of different components, thereby optimizing the fit to the surface 
brightness profile at all radii.
\end{itemize}

\begin{figure}[t]
\plotone{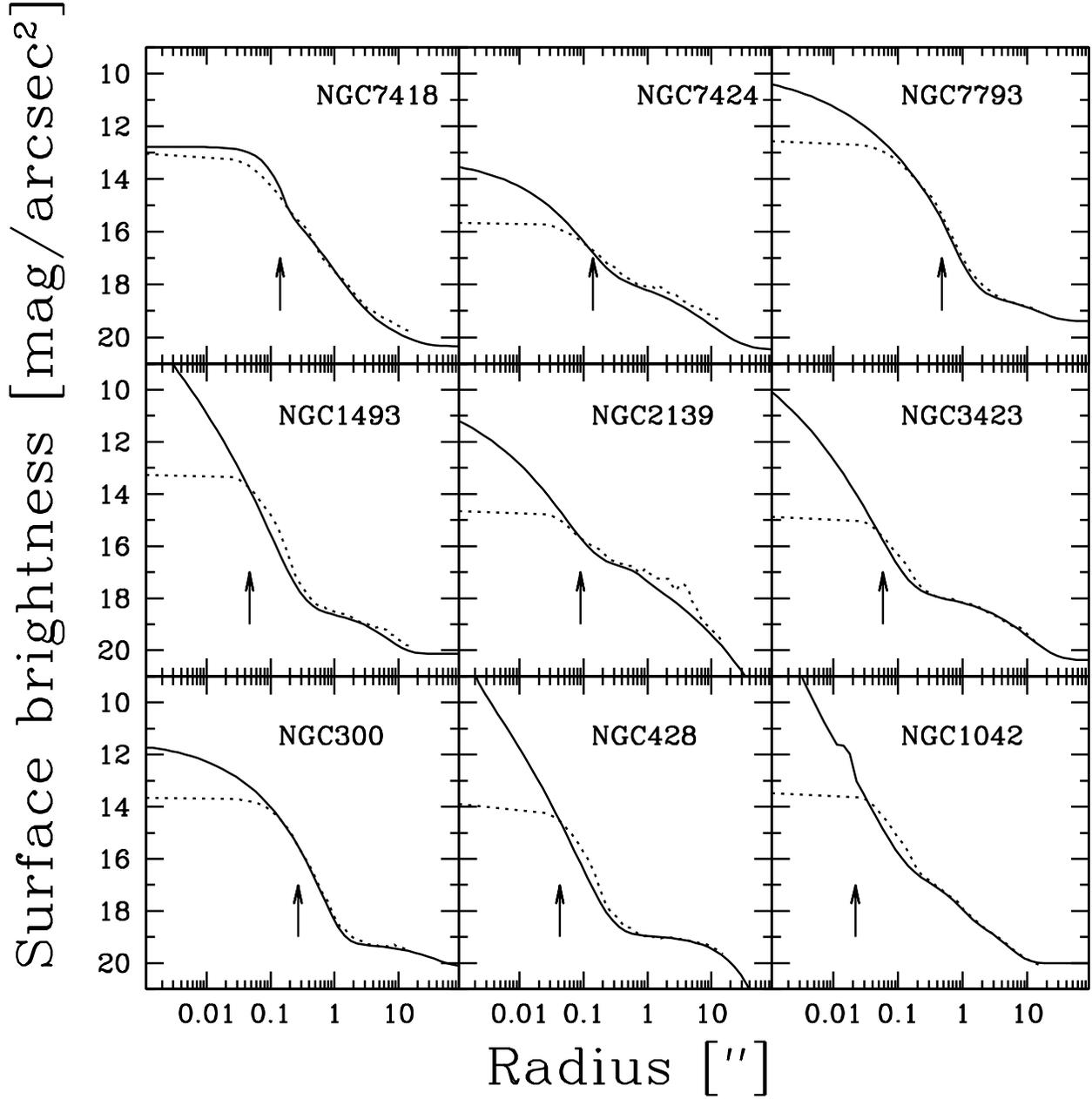}
\caption[allprof]{
The I-band surface brightness profiles 
(i.e. deconvolved from the PSF) obtained 
for the combined nuclear clusters and stellar disks in our present sample 
with {\tt galfit} (full line) compared with the non-deconvolved profiles 
obtained directly from the HST image (dotted line, from B02). 
The arrow corresponds 
to the effective radius of the star cluster. Note that 
the profile shapes agree reasonably well, but the deconvolved ones 
are always somewhat more concentrated towards the NC. 
Small differences between the
profiles at large radii are due to differences in the methods used
for model fitting and azimuthal averaging. These do not affect our
dynamical modeling results.}
\label{f:profiles}
\end{figure}

In brief, {\tt galfit} convolves a two-dimensional model which has a 
user-defined 
number of different analytic components (e.g. Sersic, Gaussian, exponential, 
constant background) with a user-supplied PSF, and finds the particular 
parameter set that best describes the observations. In most cases we use 
an exponential for the galaxy disk and a Sersic profile for the cluster. Very 
few complicated cases require more components. For example, the most 
complicated case is NGC2139, which needs five components to be fit well. 
That means a constant background, one exponential for the overall galaxy disk, 
one exponential component for the bar, one Sersic component 
for some extended emission around the cluster 
and one Sersic component for the cluster itself. 
Only one galaxy (NGC7418) shows evidence for large amounts 
of dust. Only for this galaxy we therefore created a dust mask that was used 
to cover the region in the west of the nuclear cluster from the {\tt galfit}
fit. We use the same PSFs as in B04, constructed using the 
{\tt Tiny Tim} software (Krist \& Hook 2001). Our dynamical
modeling, described in Section \ref{s:masses}, is based on the assumption of
spherical symmetry. The calculations start from the projected
surface brightness profile. For this purpose we use the
best-fitting analytic {\tt galfit} model for each galaxy to calculate the
azimuthally averaged surface brightness profile. The averaging is
done over circles that are centered on the component that we
identify as the nuclear cluster.
Figure \ref{f:profiles} shows the derived ``deconvolved'' surface brightness 
profiles in comparison to the non-deconvolved ones from B02.
While the general shape agrees well, all deconvolved profiles are more 
concentrated, as expected. Note also that the 
size of a pixel on the PC chip is 0.04555$^{\prime\prime}$. The shape of 
the profile at smaller radii is therefore not well constrained.

Although the different components of the {\tt galfit} fits were not chosen 
to have any physical meaning, it turns out that the least spatially 
extended component can be identified with the NC, except for one 
case (in NGC7418, the cluster is represented by an addition of one Sersic 
and one Gaussian profile with very similar effective radii). When comparing 
effective radii derived for the NCs under this identification with those 
determined with {\tt ISHAPE}, we find 
that they agree well to about 30\%. However strictly speaking the effective 
radii from {\tt ISHAPE} are in general more reliable than the ones derived 
from the {\tt galfit} models, because the extended wings of Sersic profiles 
may change the total luminosity content (and therefore the effective radius) 
of a model significantly with only small changes of the Sersic n parameter. 
In contrast King profiles do have a cutoff at large radius. B04 showed 
that the effective radius of the NCs is then well constrained, independent 
of uncertainties in their exact spatial profile.

%%%%%%%%%%%%%%%%%%%%%%%%%%%%%%%%%%%%%%%%%%%%%%%%%%%%%%%%%%%%%
\subsection{Dynamical Modeling}
\label{s:masses}

The luminosity $L$, effective radius $r_e$, and line-of-sight
velocity dispersion $\sigma_{\rm obs}$ are known for each of the
clusters in our sample. It follows from simple dimensional arguments
that the cluster mass-to-light ratio $M/L$ is given by
\begin{equation}
  M/L = \alpha \> \sigma_{\rm obs}^2 r_e / G L ,
\label{MLformula}
\end{equation}
where $G$ is the gravitational constant. 
The dimensionless constant $\alpha$ is determined by the density 
and velocity distribution of the cluster, and by the spatial area 
incorporated in $\sigma_{\rm obs}$. If $\sigma_{\rm obs}$ corresponds 
to the mass-weighted average dispersion of the entire cluster, then 
generally $\alpha$ is of order $\approx 10$ by virtue of the virial 
theorem. For example in this situation a spherical isotropic King 
(1962) model with concentration $c = 2$ has $\alpha = 9.77$.
The value of $\alpha$ can be adjusted to account for the fact that not all of 
the cluster stars contributed to the observed spectrum, due to the finite 
slit width, extraction aperture size, and PSF FWHM of the observations. 
However, we are dealing with clusters that reside in galaxy centers, therefore 
application of equation~(\ref{MLformula}) 
does not generally give accurate results. It ignores that some of
the light in the observed spectrum actually came from stars in the disk, 
rather than in the cluster (see the contamination fractions in Table 
\ref{t:UVESsample}). Most importantly, equation~(\ref{MLformula}) ignores the 
gravitational potential of the disk. This is important, as is most easily 
seen for stars in the outskirts of the cluster. For a model of a cluster in
isolation (such as a King model) these stars would have a dispersion
profile that falls to zero at some finite radius. However, in reality 
these stars will behave as test particles
in the gravitational potential of the disk. Their dispersion will
therefore not fall to zero, but will instead converge to the same
dispersion that the disk stars have. As a result of this effect,
equation~(\ref{MLformula}) tends to overestimate the cluster $M/L$. In
practice, we found for our sample that the results from
equation~(\ref{MLformula}) are in error by factors of up to 5. 
For accurate results it is therefore important to model the
galaxies in more detail.

We adopt an approach based on the Jeans equation for a spherical
system. The details of this approach are described in van der Marel
(1994). We start with the deconvolved galaxy surface brightness profile, 
obtained with the {\tt galfit} software as described in Section~
\ref{s:SFprof}, to which we fit a smooth parameterized profile as in Geha,
Guhathakurta \& van der Marel (2002). This profile is then deprojected
using an Abel integral under the assumption of spherical
symmetry. With an assumed constant $M/L = 1$ (in solar units) this
yields a trial three-dimensional mass density $\rho(r)$. Solution of
Poisson's equation yields the gravitational potential $\Phi(r)$, and
subsequent solution of the Jeans equation yields the three-dimensional
velocity dispersion profile $\sigma(r)$. Luminosity-weighted
projection along the line of sight gives the projected velocity
dispersion profile $\sigma_{\rm p}(R)$. Convolution with the
observational PSF and integration over pixels along the slit yields the
predicted velocity dispersion profile along the slit. Upon modeling
the extraction apertures and background subtraction procedures that
were used for the actual observations, this yields the predicted
dispersion $\sigma_{\rm pred}$. The trial mass-to-light ratio is then
adjusted to $M/L = (\sigma_{\rm obs}/\sigma_{\rm pred})^2$ to bring
the predictions into agreement with the observations. The fractional
random error on the inferred $M/L$ is twice the fractional random
error on $\sigma_{\rm obs}$. The $M/L$ values and their uncertainties
thus obtained for all the sample NCs are listed in Table~
\ref{t:masses}. The masses obtained by multiplication with the 
luminosities from B02 are also listed in the table. Errors on the 
masses include the errors on the luminosities as quoted in B02 
and the error estimate on the velocity dispersion.
In Section \ref{s:veldisp} we found that use of the Calcium Triplet 
lines for young clusters yields velocity dispersion values that are 
$\approx$ 10 km~s$^{-1}$ higher than the ones we have adopted. As 
explained there, we believe that the Calcium Triplet lines give biased 
results for young stellar populations. Nonetheless, if one were to adopt 
those results anyway, then this 
would translate into a factor of 2 
increase in mass for a cluster with a velocity dispersion of 
20 km~s$^{-1}$.

Although our results are significantly more accurate than those
obtained with equation~(\ref{MLformula}), it should be kept in mind
that there are some remaining uncertainties. First of all, each system
is assumed to be spherical. This is probably quite accurate for the
cluster, but much less so for the disk. However, this is a reasonable
lowest order approximation given that our only application is to an
extraction aperture on the center. Here there is zero net rotation,
and the disk is generally not the dominant contributor to the
observed light. Construction of fully axisymmetric models would be
more accurate, but this is considerably more complicated and beyond
the scope of the present paper. A second caveat on the adopted
approach is that we assume that the cluster and the disk have the same
$M/L$. If in reality the components have different mass-to-light
ratios then one would expect
\begin{equation}
  (M/L)_{\rm inferred} \approx (1-f) (M/L)_{\rm cluster} +
                                  f  (M/L)_{\rm disk} ,
\label{MLaverage}
\end{equation}
where $f$ is the disk light contamination fraction listed in Table~
\ref{t:UVESsample}. We used the formalism described in Geha et al. 
(2002) to calculate
several test models for one of the sample galaxies in which we assumed
different values for $(M/L)_{\rm cluster}$ and $(M/L)_{\rm disk}$. The
results confirm equation~(\ref{MLaverage}). So our inferred $M/L$
values are a weighted average of all the light in the extraction
aperture. While for some galaxies this is almost exclusively light
from the nuclear cluster, this is not true for all galaxies.

%%%%%%%%%%%%%%%%%%%%%%%%%%%%%%%%%%%%%%%%%%%%%%%%%%%%%%%%%%%%%
\section{Discussion}
\label{s:disc}

The combination of spectroscopic and photometric data enables us to 
examine the structural properties of nuclear star clusters and their 
relation to other star clusters.

%%%%%%%%%%%%%%%%%%%%%%%%%%%%%%%%%%%%%%%%%%%%%%%%%%%%%%%%%%%%%
\subsection{Comparison to other dynamically hot systems}
\label{s:plane} 

Dynamically hot stellar systems are defined 
by three observable quantities (neglecting effects of non-homology): 
velocity dispersion $\sigma_e$, 
effective radius $r_e$ and effective surface brightness $I_e$.
The total luminosity and mass are 
related to these quantities according to $L = 2 \pi r_e^2 I_e$ and 
equation (\ref{MLformula}). Dynamically hot systems satisfy a variety 
of correlations between their fundamental quantities, generally 
known under the keyword ``fundamental plane''. Any derived quantities 
will therefore also correlate. 
A popular way to view these relationships is $\kappa$ space 
(Bender, Burstein  \& Faber 1992).
The $\kappa$ parameters are defined in terms of the
fundamental observables so that $\kappa_1$ is related to mass,
$\kappa_3$ is related to $M/L$, and $\kappa_2$ is related to the
product of $M/L$ and the third power of the effective surface
brightness. Elliptical galaxies fall on a fundamental plane that
is seen edge-on when viewed as $\kappa_1$ vs.~$\kappa_3$ and is
seen face-on when viewed as $\kappa_1$ vs.~$\kappa_2$. Intrinsic 
age spreads however complicate the interpretation in $\kappa$ space, 
because $\kappa$ parameters are defined using luminosity surface 
density. Here we 
have done detailed modeling that provides mass and $M/L$ directly
of our sample NCs. We therefore choose to work with the more 
fundamental properties, rather than resort to $\kappa$ space.

In Figure 3 we plot effective projected mass density ($\Sigma_e \equiv
(M/L) I_e$) vs.~mass. This is similar, but not identical, to a
plot of $\kappa_1$ vs.~$\kappa_2$ (discussions of the latter that
are of relevance to the present topic can be found in Geha et al. 
2002 and Martini \& Ho 2004).
We plot a wide variety of different dynamical systems. 
Galaxy sized systems fill the lower right corner of the plot. Here 
small skeletal triangles show 
galaxy type systems from the compilation in Burstein et al. (1997). 
The dwarf spheroidal (dSph) galaxies have been specially 
marked with additional open 
triangles. M32, the nearest compact elliptical galaxy, is marked with an 
open pentagon filled with a star. The nucleated dE galaxies of the Virgo 
cluster from Geha et al. (2002) are represented by open squares 
filled with a cross. Stellar clusters on the other hand 
fall on a well defined band in the left of the plot. Here NCs are 
represented by black squares. The nuclei of dEs in Geha et al. (2002) 
are shown as open squares. Small stars show the locus of Milky Way 
globular clusters. 
Structural parameters ($r_e$ and total V luminosities) of 108 Galactic GCs
are derived from the online catalogue of King-model parameters maintained
by W. E. Harris (Harris 1996). Total cluster masses and related derived
quantities then follow from applying V-band mass-to-light ratios computed
by McLaughlin \& van der Marel (2004, in prep) using the
population-synthesis code of Bruzual \& Charlot (2003). 
These population-synthesis M/L values generally
compare quite well with the dynamical M/L$_V$ derived by McLaughlin (2000)
for a subsample of 40 globulars with measured velocity dispersions; see
McLaughlin \& van der Marel (2004, in prep) for more details.
The most massive globular cluster of the Milky Way, $\omega$ Cen (NGC5139), 
is specially marked as a big star. G1, the most massive globular cluster 
of M31 (Meylan et al. 2001, Baumgardt et al 2003) is also 
pointed out as a filled 
triangle. The most massive globular clusters of Centaurus A (NGC5128, Harris 
et al. 2002, Martini \& Ho 2004) are represented as starred crosses. 
Two super star clusters in M81 with dynamical mass estimates from 
McCrady, Gilbert \& Graham (2003) are shown as starred triangles. 
These were selected to be particularly bright and hence are not 
representative for the total SSC population. The young, 
peculiar globular cluster in NGC6946 described in Larsen et al. (2001) 
is denoted by an open triangle. It resides in an Sc galaxy in the disk 
at 5 kpc from the center and is surrounded by an extended star forming 
region. The cluster is extremely luminous (M$_V$=-13.2) and relatively 
massive ($1.7 \times 10^6$ M$_{\sun}$). Also shown is the most massive 
cluster known, W3 in the merger remnant galaxy NGC 7252 (Maraston et al. 2004).
We additionally plot the approximate locus of the ultra compact dwarf 
galaxies from Drinkwater et al. (2003) as a solid circle. As these authors 
do not give numbers for each one of their objects we plot only a mean 
of the range of values given in that paper. The solid line running 
through the cluster sequence represents a line of constant $r_e$ = 3 pc, 
which is appropriate for Milky Way globular clusters. 

\begin{figure}[t]
\plotone{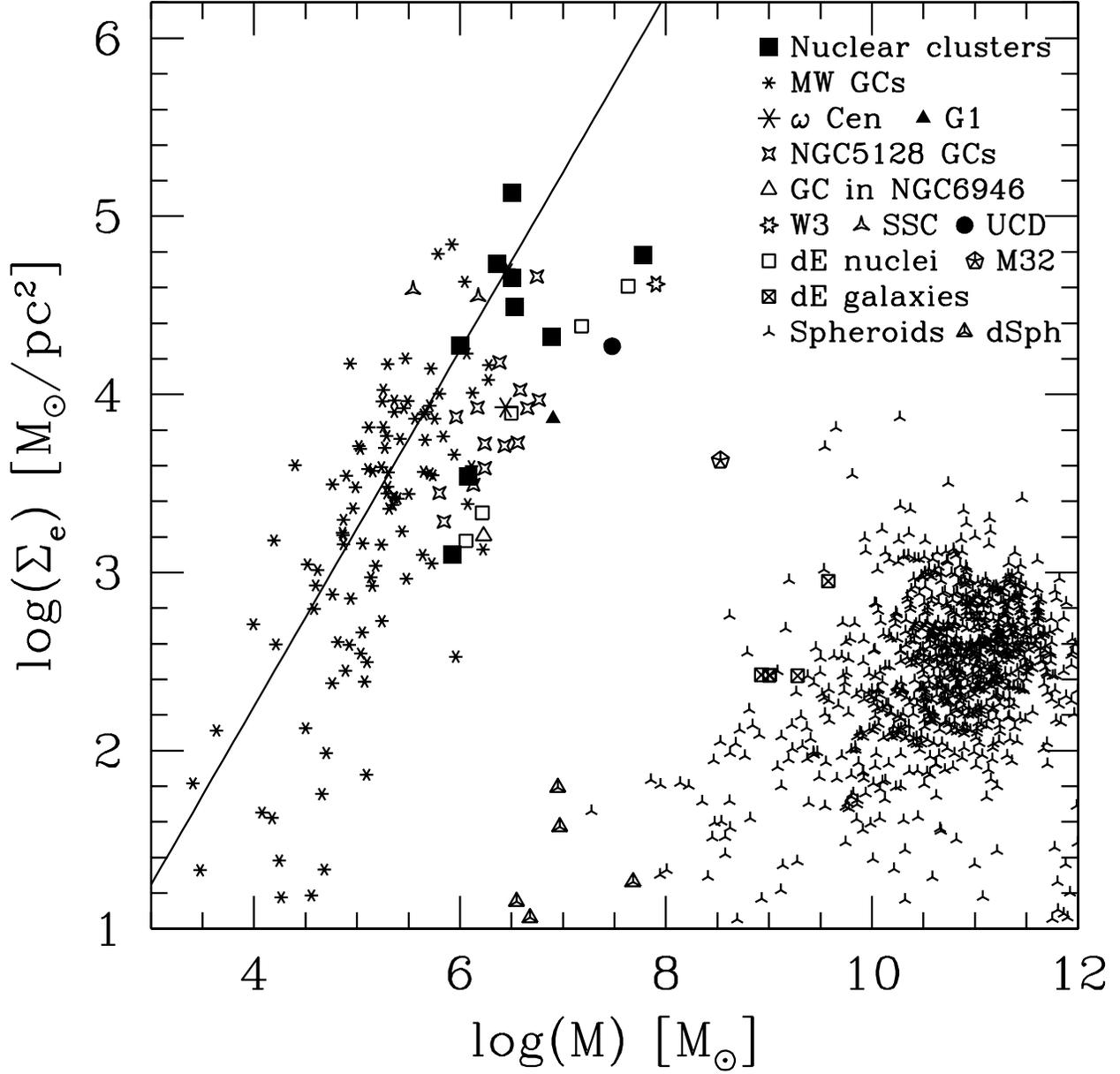}
\caption[FP face-on]{Mean projected mass density inside
the effective radius against the total mass. This is similar to 
face-on view of the fundamental plane. Symbols represent different 
types of dynamically hot stellar systems.
Nuclear clusters occupy a region together with 
different types of massive stellar clusters and are well separated from 
any bulge. The solid line represents the locus of clusters with constant 
radius, $r_e$ = 3 pc.}
\label{f:plane}
\end{figure}

Concentrating on the mass scale only, 
Figure \ref{f:plane} shows that NCs are more massive than the typical Milky 
Way globular cluster by more than one order of magnitude. While MW globular 
clusters range from $10^4$ -- $10^6$ M$_{\sun}$, NCs fall in the range $10^6$ 
-- $10^7$ M$_{\sun}$. Several extragalactic 
star clusters are however also found 
in this range. First the globular cluster system of Centaurus A extends the 
mass range of globular clusters by one order of magnitude. Then several 
clusters that have been speculated to be the remaining nuclei of accreted 
satellite galaxies are also found in this mass range (e.g. $\omega$ Cen or 
G1). Further the least massive galaxies, i.e. the dSph galaxies, are as 
massive as the most massive stellar clusters, showing that the stellar 
mass ranges of galaxies and star clusters overlap.

A clear distinction between clusters and galaxies however remains. dSph 
galaxies are less dense by 4 orders of magnitude than the most massive 
clusters. It is indeed striking that all massive clusters prolong the 
well known globular cluster sequence towards higher masses {\b and} higher 
densities. Nuclear star clusters fall well on this sequence, while there is 
a wide gap in properties to the location of compact ellipticals or bulges. 
Indeed the later Hubble type a galaxy has, the more its bulge will become 
exponential and blend with the underlying disk (compare MacArthur, Courteau 
\& Holtzman 2003 and B\"oker et al. 2003 and references therein), instead of 
remaining a small but distinct entity. Compact small bulges do not exist 
or at least have not been observed so far. While at first sight a tempting 
hypothesis, because of their 
location at the centers of their host galaxies, a smooth evolutionary 
transition from NCs to bulges therefore seems highly unlikely as there 
are no transition objects to fill the gap in Figure \ref{f:plane}.

The well-defined cluster sequence also means that, independent of their 
different environment, nuclear star clusters follow similar scaling 
relations between mass and radius as all other types of star clusters. 
From their location in Figure \ref{f:plane} the ultra compact dwarf 
galaxies of the Fornax cluster (Drinkwater et al. 2003) may well have 
to be treated as huge star clusters.
There is only a hint at a possible discontinuity in properties. While 
Milky Way globular clusters are consistent with a constant radius line, 
the cluster sequence may bend over at a characteristic scale of 
$\approx 10^6$ M$_{\sun}$, in the sense that more massive clusters 
have larger effective radii. This is analogous to the so-called 
``zone of avoidance'' in the $\kappa$-space cosmic metaplane, 
as discussed e.g. in Burstein et al. (1997). Those authors infer an 
upper limit to the 3-D luminosity density which scales roughly as 
$\sim M^{-4/3}$. In our plot of 2-D mass density, the empty upper-right 
region of Figure \ref{f:plane} suggests $\Sigma_{\rm max} \sim M^{-1/2}$.

%%%%%%%%%%%%%%%%%%%%%%%%%%%%%%%%%%%%%%%%%%%%%%%%%%%%%%%%%%%%%
\subsection{Are the centers of bulge-less galaxies
particularly conducive to Massive Cluster Formation?}
\label{s:unification}

Many authors have shown that the most massive globular clusters 
can be the remaining stripped nuclei of accreted satellite galaxies. 
In this scenario a nucleated galaxy is stripped of all of its stellar 
envelope through tidal forces during a minor merger with a bigger 
galaxy, as e.g.~our Milky Way. The nucleus however is compact enough 
to survive and remains in the halo of the bigger galaxy as a massive 
globular cluster. This scenario has been proposed for 
$\omega$ Cen (see e.g.~Bekki \& Freeman 2003 and references therein), 
G1 orbiting M31 (e.g.~Meylan et al. 2001), the ultra compact dwarf 
galaxies in the Fornax cluster (Drinkwater et al.~2003) and others.
In Figure \ref{f:plane}, the average properties of NCs are 
similar to those of the Fornax UCDs and the nuclei of dE galaxies. 
Further individual globular clusters, such as 
$\omega$ Cen and G1, have similar properties as well. 
The average properties of the Milky Way and NGC 5128 globular cluster 
sample, however are different. 
One might therefore speculate that NCs, the nuclei of dEs and UCDs 
are basically the same thing, albeit in different evolutionary stages 
of their host galaxy. Remember 
however that all of the involved samples are biased. 
The Milky Way globular cluster sample is probably not representative 
for all globular cluster systems in that it lacks the most massive ones, 
as shown by the measurements of the 
most luminous globular clusters of NGC 5128. Our own NC sample only 
covers the brighter 2/3 of the luminosity function of NCs and might 
therefore be biased to higher masses. In the case of UCDs, more 
compact and less luminous specimens might exist without being detected.

Taking Figure \ref{f:plane} at face-value however and accepting the 
merger hypothesis for the most massive clusters, it seems as though star 
clusters formed in the nuclei of galaxies are, as a class, 
more massive and more dense than the globular cluster class. 
It would then be natural to attribute these special 
properties to their location in the center of their host galaxy at 
formation time. Indeed to reach the very high space densities that 
are found in these clusters 
in one star formation event, an extremely high efficiency of star formation 
is necessary, implying high gas pressure from the surroundings of the forming 
cluster. Otherwise the gas blown out by feedback 
will invariably puff up the cluster to a much bigger size (e.g. Geyer \& 
Burkert 2001). High space densities in a quiescent environment, 
as is the case in the centers of late type spirals, can be 
reached more naturally 
by invoking repeated bursts of star formation from repeated infall of 
fresh gas. Each infalling cloud will turn some percentage of its mass 
into stars inside the boundaries of the cluster --- thus increasing the 
space density --- before blowing out the remainder through feedback. 
Rather than one single event, star formation in a nuclear star cluster 
therefore is likely to be a repetitive process.

That sufficient gas can be transported to the center of the spiral 
disk has been demonstrated by Milosavljevic (2004) under the assumption 
of a divergent dark matter density profile. 
For very late type galaxies this scenario however does not work 
well. Matthews \& Gallagher (2002) find that the rotation 
curves of very late type spirals are nearly linear, implying a 
constant surface mass density and a harmonic potential, where 
the gravitational vector vanishes at the center. It therefore remains 
unclear, why the centers of bulge-less galaxies would be special 
places, conducive to the formation of stellar systems with 
extreme physical properties.

%%%%%%%%%%%%%%%%%%%%%%%%%%%%%%%%%%%%%%%%%%%%%%%%%%%%%%%%%%%%%
\subsection{Phase space densities}
\label{s:density}

\begin{figure}[t]
\plotone{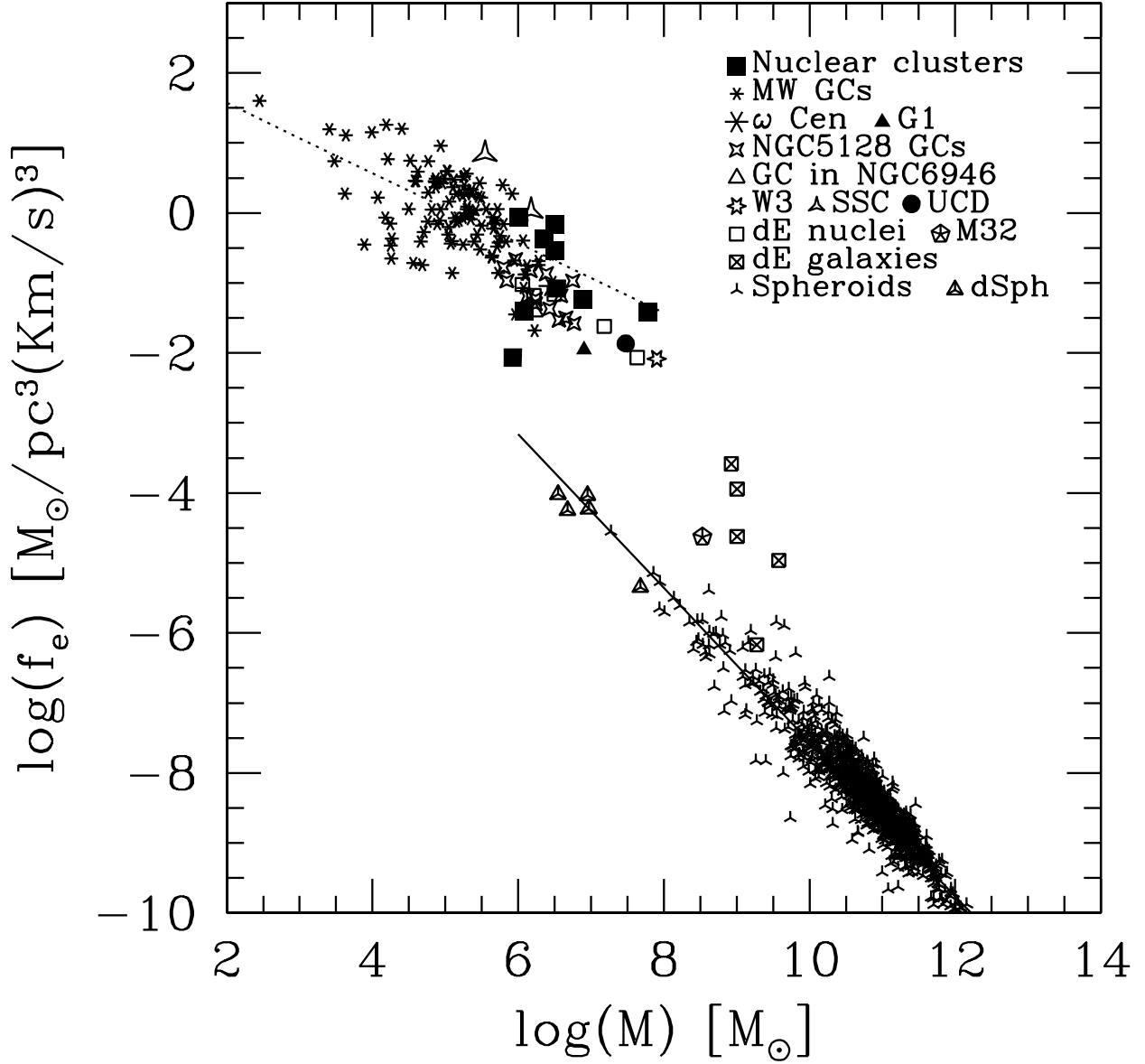}
\caption[compactness]{The characteristic phase space density $f_h$, i.e. 
the mean phase space density inside the half-mass radius $r_h$ for 
the same dynamically hot systems as in Figure \ref{f:plane}. The dotted 
line represents the locus of systems that obey the virial theorem and 
have constant radius $r_h$. The solid line represents a
locus of systems that obey the virial theorem as well as a
Faber-Jackson (1976) type relationship of the form $M \propto
\sigma^{3.33}$. }
\label{f:effdens}
\end{figure}

Figure \ref{f:effdens} shows $f_h$, the characteristic phase space 
density inside the half-mass radius, against mass for the same types of 
stellar systems as used in Figure \ref{f:plane}. We define $f_h$ using 
the half-mass radius $r_h$, total mass $M$ and measured velocity 
dispersion $\sigma$ according to 
\begin{equation}
  f_h = \frac{\rho_h}{\sigma^3} = \frac{M}{2} \frac{1}{\frac{4}{3}\pi r_h^3 \sigma^3} \propto r_h^{-2} \sigma^{-1}.
\end{equation}
This quantity can be calculated for various dynamically hot systems 
using the literature mentioned in Section \ref{s:plane}.
The projected effective radii $r_e$ have been converted to 3-D half-mass 
radii $r_h$ by the approximate relation $r_e$ = 0.75 $r_h$ (see Spitzer 
1987, p. 12). No correction was applied to the velocity dispersions.

Additionally the lines in Figure \ref{f:effdens} show fits to two 
different subsets of the systems. The dotted line is a fit to Milky 
Way globular clusters only, where $\log(f_h) = 2.57 - 0.5 * \log(M)$. 
The virial theorem dictates that $M \propto r_h \sigma^2$. The fit 
therefore is a line of constant $r_h$, where $r_h \approx 8$ pc. 
The solid line is a fit to the 
galaxy type systems, where $\log(f_h) = 3.44 - 1.1 * \log(M)$ 
(excluding the Virgo dE,N from Geha et al. 2002).
This fit implies that $M \propto \sigma^{3.3}$, which 
is a  Faber-Jackson (1976) type relationship.

Again as in Section \ref{s:plane} nuclear clusters fall on the 
locus of typical massive star clusters. There is a clear discontinuity to 
galaxies, but more importantly a change of slope, thereby reinforcing the 
statement that nuclear star clusters are typical massive star clusters 
(and not progenitors of bulges). 

On the other hand the phase space densities may be directly linked 
to star formation processes. Dynamical evolution is potentially 
not responsible 
for the slope observed in the cluster sequence as there is no obvious 
trend with age, even though the ages of the 
involved clusters range from around $10^7$ years (SSCs, YMC in NGC6946) over 
$10^8$ years (some of the NCs, W3) to very old objects like the 
Milky Way globular clusters. A more detailed interpretation 
of this result clearly requires simulations of the formation of 
massive star clusters and NCs in particular.

%%%%%%%%%%%%%%%%%%%%%%%%%%%%%%%%%%%%%%%%%%%%%%%%%%%%%%%%%%%%%
\subsection{Formation of intermediate mass black holes}
\label{s:IMBH}

Due to their very high mass densities, NCs might be considered 
to be ideal candidates for fast core collapse and subsequent runaway 
merging of young massive stars, thus forming an intermediate mass 
black hole. Comparison with relevant current models (Portegies-Zwart et al. 
2004, Marc Freitag, priv. comm.) however shows that NCs do not fall into 
the appropriate region of parameter space. When inserting the typical 
values of NCs into equation (1) of Portegies-Zwart et al. (2004) we 
obtain a dynamical friction time scale of the order of 24 Myr for a 
100 M$_{\sun}$ star, which is much longer than its lifetime. These 
stars therefore will explode as supernovae before reaching the center 
of the cluster and will not experience runaway merging.
Less-massive stars undergo weaker dynamical friction, and are thus 
even less likely to reach the cluster center in an appropriate 
timescale for the formation of a black hole.

%%%%%%%%%%%%%%%%%%%%%%%%%%%%%%%%%%%%%%%%%%%%%%%%%%%%%%%%%%%%%
\section{Summary and conclusions}
\label{s:concl} 

Most bulge-less, very late-type galaxies have been shown to have 
photometrically distinct, compact nuclear star clusters (several pc in 
size) at their photometric centers, a stark contrast to the diffuse 
stellar disk around them. We determine the stellar velocity dispersion 
for 9 such nuclear star clusters, derived from high resolution spectra 
taken with UVES at the VLT. For observational reasons we 
sample the brighter 2/3 of the 
luminosity range covered by the clusters. 
In conjunction with light profiles from 
the WFPC2 camera on board the HST, we also determine their masses. 
We find them to range from $8\times 10^5$ M$_{\sun}$ to $6\times 
10^7$ M$_{\sun}$. The nine objects analyzed provide an order of 
magnitude increase in the number of available determinations, 
as the only mass known so 
far had been determined in IC342 to be $6\times 10^6$ M$_{\sun}$ 
(B\"oker et al. 1999). 

The mass estimates show that as a class, these nuclear 
clusters in late type galaxies have structural properties similar 
to the most massive known stellar clusters.
 
\begin{itemize}
\item At around $5 \times 10^6$ M$_{\sun}$ their characteristic masses are 
much larger than those of Milky Way globular clusters, except $\omega$ Cen.
They however fall in the same range as some of the most extreme stellar 
clusters observed so far, as e.g. G1 in M31 or the most massive 
globular clusters of NGC 5128.

\item The properties of the nuclear clusters 
show that they are widely distinct from all bulges with 
measured structural parameters, by orders of magnitude in mass and radius.

\item Combining the mass estimates with their effective radii, we find 
nuclear clusters to be among the clearly distinct stellar systems with
the highest mean mass density within their effective radius. Remarkably, 
although they 
lie in hugely different environments and thus presumably form in different 
ways, different sorts of star clusters follow the same mass to density and 
mass to phase-space density relations.
\end{itemize}

The combination of a galaxy center and a dynamically quiescent environment
appears to be conducive to the creation of dense stellar systems with 
extreme physical properties. 
These results thus confirm 
that even in these very late-type galaxies, of overall low stellar surface 
mass density, the center of the galaxy is 'special'. This leaves open 
the question, of why some late-type galaxies seem 
not to have any such clusters.
From their position in their host galaxies, the closest relatives of 
nuclear clusters appear to be the nuclei of 
dwarf ellipticals, which could be older and somewhat more diffuse analogs. 
However, assuming that massive globular clusters and the Fornax 
ultra-compact dwarf galaxies are the remainders of accreted nucleated 
dwarf galaxies, all massive clusters with very high effective 
mass densities could have a common origin in the centers of galaxies.

In a companion paper (Walcher et al. 2004 in prep.) we will present further 
results from the blue parts of our UVES spectra yielding ages and star 
formation rates in our sample of nine nuclear clusters. We already remark 
here that the I-band mass-to-light ratios we find for nuclear clusters are 
smaller than the typical value of 1.2 (McLaughlin 2000) found for 
Galactic globular clusters and thus point to significantly younger 
objects.

%%%%%%%%%%%%%%%%%%%%%%%%%%%%%%%%%%%%%%%%%%%%%%%%%%%%%%%%%%%%%
\section*{ACKNOWLEDGMENTS}       

CJW would like to acknowledge intensive discussions with Marla Geha 
and the help of Dan Choon with literature research. CJW also thanks 
STScI for hosting a long term visit under a directors 
discretionary research fund grant. This research has 
made use of the SIMBAD astronomical database, created and maintained by 
the CDS, Strasbourg, as well as NASA's Astrophysics Data System and of 
the NASA/IPAC Extragalactic Database (NED) which is operated by the 
Jet Propulsion Laboratory, California Institute of Technology, under 
contract with the National Aeronautics and Space Administration.
Based on observations collected at the European Southern Observatory, 
Chile (ESO Programme 68.B-0076(A)).

%%%%%%%%%%%%%%%%%%%%%%%%%%%%%%%%%%%%%%%%%%%%%%%%%%%%%%%
%%%%% References %%%%%

\clearpage

\clearpage

\clearpage

\clearpage

\clearpage

\begin{table}[t]
\caption{Sample and observations}
\begin{tabular}{ccccccc}
\tableline\tableline 
{Galaxy}  & {type} & {Distance}& {m$_I$ }& {M$_{I}$} & {S/N}    & {Contamination}\\
          &        & {[Mpc]}   &         &           &          &    {[\%]}     \\
    (1)   &  (2)   &     (3)   &   (4)   &    (5)    &   (6)    &     (7)      \\
\tableline
NGC~300   & SAd    & 2.2       & 15.29   &    -11.43 &  37      &  0 \\
NGC~428   & SABm   & 16.1      & 17.95   &    -13.15 &  16      & 30 \\
NGC~1042  & SABcd  & 18.2      & 18.40   &    -12.95 &  24      & 81 \\
NGC~1493  & SBcd   & 11.4      & 17.17   &    -11.43 &  20      & 33 \\
NGC~2139  & SABcd  & 23.6      & 19.28   &    -12.65 &  19      & 89 \\
NGC~3423  & SAcd   & 14.6      & 19.04   &    -11.84 &  15      & 76 \\
NGC~7418  & SABcd  & 18.4      & 15.12   &    -16.23 &  41      & 18 \\
NGC~7424  & SABcd  & 10.9      & 18.80   &    -11.41 &  16      & 77 \\
NGC~7793  & SAd    & 3.3       & 14.00   &    -13.64 &  65      &  0 \\
\tableline
\end{tabular}
\label{t:UVESsample}
\tablecomments{Cols. (1) and (2) Galaxy name and type as taken from NED. 
Col. (3) Distances were taken from B02, where they were calculated 
from the recession velocity (from LEDA, 
corrected for virgocentric infall) and assume 
H$_0$ = 70 km~s$^{-1}$. Col. (4) and (5) Apparent and absolute magnitude 
of the NC as taken from B\"oker et al. (2002). Col. (6) Signal-to-noise 
per pixel ($\sim 0.07$ \AA) in the region around the Calcium Triplet. 
Col. (7) Contamination from galaxy disk light in percent, measured as 
described in Section \ref{s:reduction}.}
\end{table}

\begin{table}[t]
  \caption{Stellar template weights}
    \begin{tabular}{c|cccccccccccc}
     \tableline \tableline
     {Star}      & {type} & {m$_V$} & 300 & 428 & 1042 & 1493 & 2139 & 3423 & 7418 & 7424 & 7793 \\
     \tableline
     HR3611      & B6V    & 5.60 & 0.0  & 0.0  & 0.0  & 0.0  & 0.0  & 0.0  & 0.0  & 0.0  & 0.32\\
     HR3230      & A1V    & 6.52 & 0.26 & 0.31 & 0.63 & 0.75 & 0.76 & 0.36 & 0.79 & 0.73 & 0.19\\
     HR3473      & A5V    & 6.11 & 0.12 & 0.0  & 0.0  & 0.0  & 0.0  & 0.0  & 0.0  & 0.0  & 0.0\\
     HR3070      & F1V    & 5.78 & 0.0  & 0.0  & 0.0  & 0.0  & 0.0  & 0.0  & 0.0  & 0.0  & 0.0\\
     HR3378      & G5III  & 5.87 & 0.0  & 0.0  & 0.0  & 0.0  & 0.0  & 0.0  & 0.0  & 0.0  & 0.0\\
     HR1167      & G8III  & 6.49 & 0.0  & 0.59 & 0.25 & 0.18 & 0.22 & 0.50 & 0.8  & 0.23 & 0.48\\
     HR1216      & K2III  & 5.93 & 0.61 & 0.09 & 0.12 & 0.07 & 0.02 & 0.13 & 0.13 & 0.4  & 0.01\\
     \tableline
    \end{tabular}
   \label{t:UVESstars}
\tablecomments{Cols. (4) to (12) are labelled by NGC number and 
contain the relative weights of the template stars that contribute 
to the velocity dispersion fit, as 
determined by the fitting code.}
\end{table}

\clearpage

\begin{table}[t]
  \caption{Dynamical quantities} 
    \begin{tabular}{cccccc}
     \tableline \tableline 
     {Galaxy}    &  {$\sigma$ [km~s$^{-1}$]} & {log(L$_I$) [L$_{\sun}$]}  & {M/L$_I^{dyn}$} & {log(M) [M$_{\sun}$]}\\
    (1)            &  (2)           &     (3)         &   (4)         &    (5)    \\
     \tableline
     NGC~300       & $13.3 \pm 2.0$ &  $6.21\pm0.20$  &  $0.65\pm0.20 $  & $6.02 \pm 0.24 $ \\
     NGC~428       & $24.4 \pm 3.7$ &  $6.89\pm0.04$  &  $0.42\pm0.13 $  & $6.51 \pm 0.14 $ \\
     NGC~1042      & $32.0 \pm 4.8$ &  $6.81\pm0.16$  &  $0.50\pm0.15 $  & $6.51 \pm 0.21 $ \\
     NGC~1493      & $25.0 \pm 3.8$ &  $6.89\pm0.05$  &  $0.31\pm0.09 $  & $6.38 \pm 0.14 $ \\
     NGC~2139      & $16.5 \pm 2.5$ &  $6.69\pm0.16$  &  $0.17\pm0.05 $  & $5.92 \pm 0.20 $ \\
     NGC~3423      & $30.4 \pm 4.6$ &  $6.37\pm0.06$  &  $1.46\pm0.44 $  & $6.53 \pm 0.14 $ \\
     NGC~7418      & $34.1 \pm 5.1$ &  $8.13\pm0.13$  &  $0.45\pm0.14 $  & $7.78 \pm 0.19 $ \\
     NGC~7424      & $15.6 \pm 2.3$ &  $6.20\pm0.06$  &  $0.78\pm0.23 $  & $6.09 \pm 0.14 $ \\
     NGC~7793      & $24.6 \pm 3.7$ &  $7.08\pm0.05$  &  $0.64\pm0.19 $  & $6.89 \pm 0.14 $ \\
     \tableline
    \end{tabular}
   \label{t:masses}
\tablecomments{Col. (1) Galaxy name. Col. (2) Measured velocity dispersion 
with systematic error. Col. (3) Logarithm of the I-band cluster luminosity in 
L$_{\sun}$ (from B02). The errors we use here are somewhat larger than 
quoted in 
B02 because they include the uncertainty of 0.1 mag due to difficulties 
in determining the outer radius of the cluster. 
Col. (4) and (5) $M/L$ (in solar I-band units) 
and $\log(M) = \log(M/L * L)$ of the cluster as derived from the dynamical modeling.}
\end{table}

\end{document}